\documentclass[]{raa}
\usepackage{graphicx,times}
\usepackage{natbib}
\usepackage{indentfirst}
\usepackage{ifpdf}

\providecommand{\bm}[1]{\mbox{\boldmath$#1$\unboldmath}}

\begin{document}
\title{The Broad Band Spectral Energy Distributions of SDSS Blazars}
\volnopage{Vol.0 (200x) No.0, 000--000}
\date
   \setcounter{page}{1}
   \author{Huaizhen Li \inst{1},Luoen Chen \inst{1}, Yunguo Jiang\inst{2} and Tingfeng Yi \inst{3}}
   \institute{Physics Department, Yuxi Normal University, Yuxi 653100, China
           \and
            Shandong Provincial Key Laboratory of Optical Astronomy and Solar-Terrestrial Environment, Institute of Space Sciences, Shandong University,Weihai, 264209, China; {\it jiangyg@sdu.edu.cn}\\
            \and
             Physics Department, Yunan Normal University, Kunming 650092, China\\
        \vs \no
   {\small Received ********; accepted ********}}

\abstract{We compiled the radio, optical, and X-ray data of blazars from the
Sloan Digital Sky Survey (SDSS) database, and presented the
distribution of luminosities and broad band spectral indices. The
distribution of luminosities shows that the averaged luminosity of flat-spectral
radio quasars (FSRQs) is larger than that of BL Lacs objects.
On the other hand, the broad band spectral energy distribution reveals that
FSRQs and low energy peaked BL Lac objects (LBLs) objects have
similar spectral properties, but high energy peaked BL Lac objects
(HBLs) have a distinct spectral property.
This may be due to that different subclasses of blazars have different intrinsic environments and are at different cooling levels.
Even so, a unified scheme also is revealed from the color-color
diagram, which hints that there are similar physical processes
operating in all objects under a range of intrinsic physical
conditions or beaming parameter.
 }
   \authorrunning{H. Z. Li et al.}
   \titlerunning{The Broad Band Spectral Energy Distributions of SDSS Blazar}
   \maketitle

\section{Introduction}
Blazars are a subset of active galactic nuclei (AGNs) with strong emission at all wavelengths. They are the
brightest and most variable high energy sources among AGNs, and have continuous spectral energy distributions (SEDs). The continuum emission in balzars is thought to be from a
relativistic jet oriented close to the observer and emanating from the
vicinity of a black hole Ster08,Ghis86. The SEDs of blazars
are characterized by a universal two-bump structure: one in the IR
to UV band, another in the MeV-GeV band. The synchrotron radiation
in a relativistic beamed jet is responsible for the lower-energy peak,
while the high-energy $\gamma$-ray emission is produced by the
inverse Compton mechanism \citep{Samb96}. Generally, blazars can be divided into two subclasses, the BL
Lacertae (BL Lacs) object and the flat spectrum radio quasars (FSRQs).
The main difference between these two classes is their
emission lines: BL Lacs objects are characterized by the lack of strong
emission lines (equivalent width $<$ 5${\AA}$, while FSRQs have
strong broad emission lines with the similar strength of normal quasars
\citep{Scar97}.

Based on the synchrotron emission peak energy frequency, BL Lacs objects can be divided into "High energy peaked BL Lacs" (HBLs) and
"Low energy peaked BL Lacs" (LBLs) \citep{Pado95,Giom95}.
\citet{Mei02} and  \citet{Ma07} found that the two subclasses of BL
Lacs objects can be distinguished by the peak frequencies $\nu_{peak}$: HBLs have $log\nu_{peak}>14.7$,
 while LBLs have $log\nu_{peak}<14.7$. In addition, \citet{Pado95} found that LBLs and HBLs can also be divided by using the ratio of X-ray flux at 1 keV (in units of $erg$ $cm^{-2}$ $s^{-1}$) to 5 GHz radio flux density (in units of Janskys). The criterion is $f_{x}/f_{r}\sim10^{-11}$, corresponding to the broad
band spectral index (from radio 5 GHz to X-ray 1 keV) $\alpha_{rx}\simeq0.75$.
HBLs have the broad band spectral index of $\alpha_{rx}\leq0.75$, and LBLs have the spectral index of $\alpha_{rx}> 0.75$ \citep{Giom95,Ma07,Mei02,Urry95}. For two different subclasses of BL
Lacs objects, the SEDs have been investigated by a number of authors
\citep[e.g.,][]{Bao08,Chen06,Pado95,Giom90,Niep06}. \citet{Pado95}
and \citet{Giom90} found that two subclasses occupy different regions on
the $\alpha_{ro}$ - $\alpha_{ox}$ plane. \citet{Pado95} also found
there exist the correlations between the minimum soft X-ray flux and
the radio flux, and also the correlations between radio and optical fluxes for the
subsample of HBLs, but not for that of LBLs. \citet{Niep06} found
that there is a negative correlation between the luminosity and
the synchrotron peak frequency $\nu_{peak}$ at radio and optical band,
whereas the correlation turns slightly positive in X-ray
\citep{Niep06}. \citet{Fan12} and \citet{Lyu14} found that HBLs have different properties from LBLs. \citet{Yan14} found that the one-zone synchrotron self-Compton (SSC) model can successfully fit the SEDs of HBLs, but fails to explain the SEDs of LBLs. In addition, \citet{Bao08} found that two subclasses of BL Lacs objects are unified.

BL Lacs objects and FSRQs are grouped together under the
denomination of blazars, which eliminates the somewhat ambiguous
issue of strength of the emission lines as a classification
criterion. However, there are some differences in the individual
emission properties among different blazar subclasses. The
relationship among different kinds of balzars can promote our
perception of the fundamental properties of blazars. Therefore, it
is imperative to investigate the connection among FSRQs, LBLs, and
HBLs.

The relationship between of BL Lacs objects and FSRQs has been
discussed by a number of authors
\citep[e.g.,][]{Coma97,Foss98,Ghis98,Ghis09,Li10,Samb96,Xie01,Xie04a,Xie04b,Xie06,Xie07,Xie08,Zhen07},
who assembled the SED of many radio, X-ray, and $\gamma$-ray
selected blazars. \citet{Foss98} studied the SEDs of a combined
blazar sample, and found that the SEDs properties of these
subclasses present a remarkable continuity and a systematic trend as
a function of source luminosity, which suggests that the parameter
describing the blazar continua is likely to be the source
luminosity. Based on the first Fermi sample, \citet{Ghis09} found
that FSRQs and BL Lacs objects occupy separate regions, and obey a
spectral sequence. However, \citet{Ant05} found that there are selection
effects for the "blazars sequence" reported by \citet{Foss98} and
\citet{Ghis98}. Some literatures show that HBLs have different properties from FSRQs, but LBLs are similar to FSRQs \citep[e.g.][]{Chen13,Fan12,Li10,Lyu14}.
However, \citet{Li10} and \citet{Chen13} also found that their whole sample suggest the unified
scheme of blazars. \citet{Coma97} discovered
that there is a significant anticorrelation between X-ray and
$\gamma$-ray spectral indices, and also between the broadband spectral
indices $\alpha_{ro}$ and $\alpha_{x\gamma}$ of BL Lacs objects and
FSRQs. The correlation between the broadband spectral indices
obtained by \citet{Coma97} implied that there is a different shape in
overall energy distributions from radio to $\gamma$-ray energies
between BL Lacs objects and FSRQs. \citet{Samb96} and \citet{Xie04a}
found that three kinds of blazars have different SEDs, but follow a
continuous spectral sequence.

In this paper, we will study the distributions of luminosities and the
radio-optical-X-ray SEDs of SDSS blazars, and connections among
LBLs, HBLs, and FSRQs. A detailed explanation of the sample is given
in Sec. 2. The distributions of luminosity are presented in Sec. 3.
The broad band spectral energy distribution is given in Sec. 4. In Sec
5, discussions and conclusions are presented.

\section{The Sample of SDSS Blazars}
The Sloan Digital Sky Survey (SDSS) is one of the most ambitious and
influential surveys in the history of astronomy. \citet{Plot08} have
drawn a large sample of 501 BL Lacs objects candidates from the
combination of SDSS Data Release 5 (SDSS DR5) optical spectroscopy, and the
Faint Images of the Radio Sky at Twenty-Centimeters (FIRST) radio
survey. \citet{Plot10} have presented a sample of 723 optically selected BL Lac candidates from the SDSS Data Release
7 (SDSS DR7) spectroscopic database. Based on the large radio (the NRAO VLA Sky Survey, ATCA catalogue of compact PMN sources), ROSAT All Sky Survey (RASS), the SDSS Data Release 4 (SDSS DR4) and 2dF survey data, \citet{Turr07} presented a Radio-Optical-X-ray catalog built at ASDC (ROXA) including 816 objects, among which 510 are confirmed blazars. In addition, \citet{Chen09} have also presented a
sample including 118 Non-thermal jet-dominated FSRQs from SDSS Data Release 3 (SDSS DR3), X-ray quasar sample with
FIRST and GB6 radio catalogues. Based on the sample of
\citet{Plot08,Plot10}, \citet{Turr07} and \citet{Chen09}, we compiled a large
sample of 606 blazars, including 292 FSRQs and 314 BL Lacs. All the objects of the sample have matches in RASS and measured redshifts.
In our sample, the three-band luminosities
($L_{r}$, $L_{o}$, and $L_{x}$) and the broad-band spectral indices
$\alpha_{ro}$, $\alpha_{rx}$, and $\alpha_{ox}$ were
given by the literatrues \citep{Chen09,Plot08,Plot10,Turr07}.
The luminosities $L_{r}$, $L_{o}$, and $L_{x}$ are the specific
luminosities (per unit frequency) at 5GHz, 5000${\AA}$, and 1keV,
respectively. $\alpha_{ro}$ is the two-point spectral indices
between 5 GHz and 5000${\AA}$, $\alpha_{rx}$ is the two-point
spectral indices between 5 GHz and 1 keV, and $\alpha_{ox}$ is the
two-point spectral indices between 5000${\AA}$ and 1 keV.

As discussed in section 1, BL Lacs objects can be divided into HBLs
and LBLs, based on the radio-X-ray spectral index $\alpha_{rx}$
between 5 GHz and 1 keV. According to literatures
\citep{Giom95,Ma07,Mei02,Plot08,Urry95}, most BL Lacs objects with
$\alpha_{rx}\leq0.75$ are HBLs, and most BL Lacs objects with
$\alpha_{rx}>0.75$ are LBLs. Therefore, for investigating the
relation among different blazar subclasses, we adopt
$\alpha_{rx}=0.75$ as a rough value to divise HBLs and LBLs for the
SDSS BL Lacs objects \citep{Giom95,Ma07,Mei02,Urry95}. Based on this
criterion, there are 270 HBLs and 44 LBLs in our sample.

\section{Distributions of luminosity of blazars}
\citet{Foss98} studied the SEDs of a combined blazar sample, and
found that the source luminosity is the characteristic parameter describing the blazar
continua. On the basis of the first Fermi sample, \citet{Ghis09}
have found BL Lacs objects are harder and less luminous than FSRQs.
\citet{Ghis98} found that HBLs are sources with the lowest intrinsic
power and the weakest external radiation field, LBLs are
intrinsically more powerful than HBLs, and FSRQs represent the most
powerful blazars.

Thus, we computed the distributions of radio (at 5GHz), optical (at
5000${\AA}$) and X-ray (at 1 keV) luminosities for three subclasses
of blazars. Figure 1-3 give the distribution of luminositices for
three kinds of blazars, and all the luminositices are K-corrected to the
source rest frame \citep{Chen09,Plot08,Plot10,Turr07}. The distributions of radio,
optical and X-ray luminosities are plotted in Figure 1, 2, and 3,
respectively.

From Figure 1, one can find that FSRQs have larger radio
luminosities than BL Lacs objects, while the radio luminosities of
LBLs are more powerful than that of HBLs. This suggests that the
radio luminosities of the three kinds of blazars, from FSRQs to LBLs
to HBLs, are decreasing, which is consistent with the argument
reported by \citet{Ghis98}. In Figure 2, we can note that the
optical luminosities of FSRQs are larger than that of BL Lacs
objects, while HBLs have similar optical luminosities to that of
LBLs. Correspondingly, the X-ray luminosities of HBLs are systematically lower than
FSRQs, but larger than LBLs (see Figure 3).

The distributions presented in Figure 1, 2, and 3 show that the
luminosity is an important parameter to distinguish FSRQs and BL
Lacs objects. A tendency of luminosities from FSRQs to BL Lacs objects is revealed from the distributions of
luminosities. The distributions of luminosities for three kinds of
blazars presented from Figure 1-3 are consistent with the results
reported by \citet{Foss98} and \citet{Ghis98}. On the other hand,
one can note that all the distributions are continuous in properties
between HBLs and LBLs, and as well as between FSRQs and BL Lacs
objects, which is in good agreement with the previous arguments
about the continuum of blazars \citep{Foss98,Ghis98,Xie04a,Coma97,Samb96}.

\begin{figure}
\includegraphics  [width=5in, angle=0]{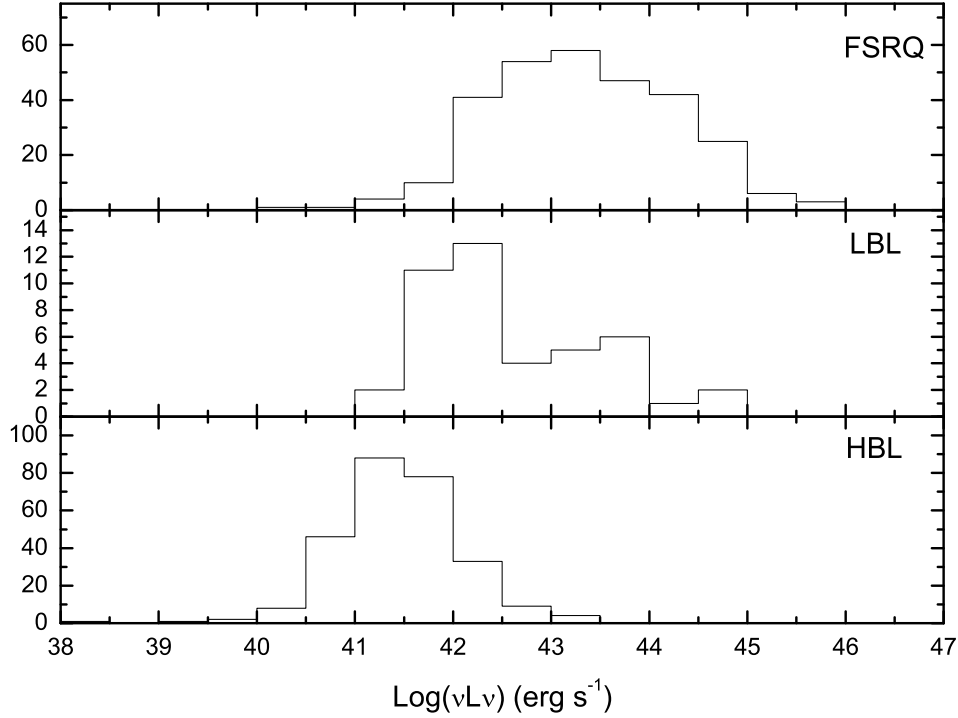}
 \caption{Distributions of radio luminosity at 5 GHz for three kinds of blazars of our sample.
\label{fig1}}
\end{figure}


\begin{figure}
\includegraphics  [width=5in, angle=0]{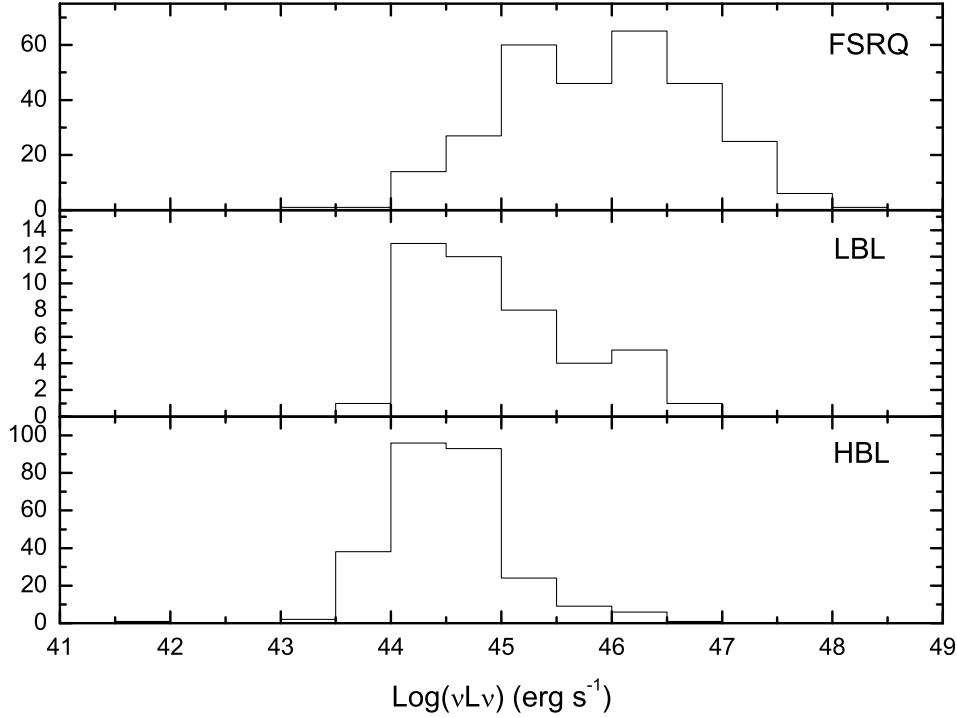}
 \caption{Distributions of optical luminosity at 5000${\AA}$ for three kinds of blazars of our sample.
\label{fig2}}
\end{figure}


\begin{figure}
\includegraphics  [width=5in, angle=0]{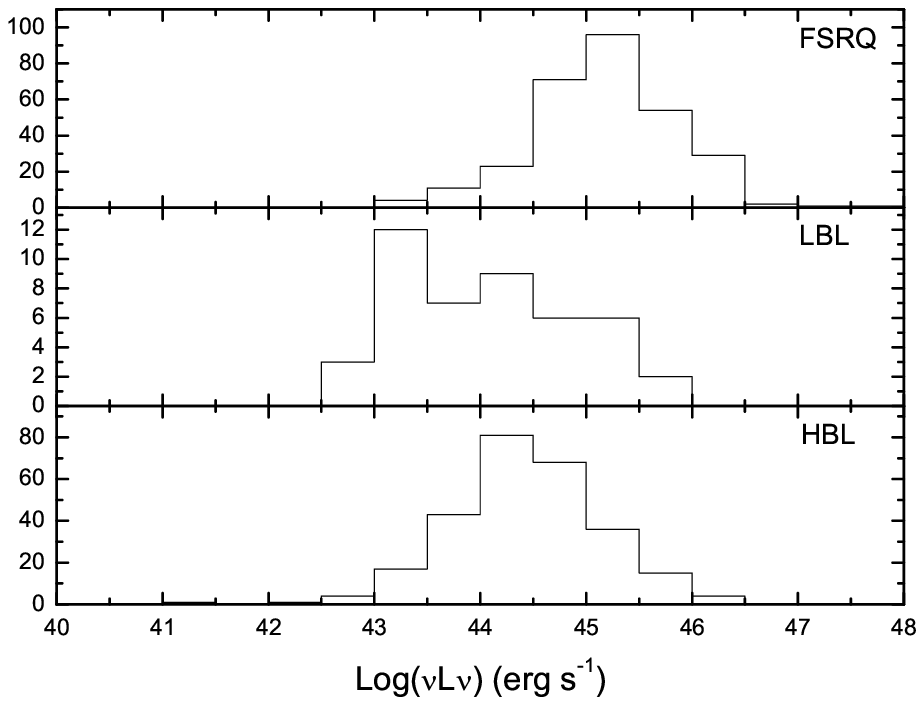}
 \caption{Distributions of X-ray luminosity at 1 keV for three kinds of blazars of our sample.
 \label{fig3}}
\end{figure}

\section{Broad band spectral energy distribution of blazars}
Searching the connection among different blazar subclasses is very
significant, because it can substantially promote our understanding
about the fundamental nature of blazars. The relationship among
different blazars has been discussed in literatures with a
unified scheme and a spectral sequence for blazars
\citep{Foss98,Ghis98,Samb96,Xie04a}. For investigating the
relationship among different subclasses of blazars, we will analyze
the relationship among the HBLs,
 LBLs, and FSRQs on the basis of the broad band spectral index
$\alpha_{ro}$, $\alpha_{rx}$, and $\alpha_{ox}$.

\subsection{Diagram of $\alpha_{rx}$ versus $\alpha_{ro}$}

Here, we investigated the relationship between the broad band spectral indices $\alpha_{rx}$ and $\alpha_{ro}$ for the whole sample. The plot is shown in Figure 4.
Figure 4 shows that there is a
good correlation between $\alpha_{rx}$ and $\alpha_{ro}$ for the whole sample. A linear
regression analysis equation for all sample is written as
\begin{equation}
\alpha_{rx}=(0.64\pm 0.03)\alpha_{ro}+(0.43\pm0.01),
\label{eq:Lebseque1}
\end{equation}
with a correlation coefficient $r=0.70$, and a chance
probability $p<10^{-4}$.
Moreover, we also studied the relationship between $\alpha_{rx}$ and $\alpha_{ro}$ for the FSRQs and LBLs sample. We obtained

\begin{equation}
\alpha_{rx}=(0.42\pm 0.02)\alpha_{ro}+(0.59\pm0.01),
\label{eq:Lebseque2}
\end{equation}
with a correlation coefficient $r=0.70$, and a chance
probability $p<10^{-4}$. The correlation analysis suggests that there is a linear correlation between $\alpha_{rx}$ and $\alpha_{ro}$ for the whole sample, and as well as for the FSRQs+LBLs sample. This provides evidence for the
unified scheme reported by \citet{Foss98} and \citet{Ghis98}.

In addition, Figure 4 shows that the majority of FSRQs and LBLs mix
together, which suggests that they have similar spectral properties.
However, Figure 4 also reveals that the distribution of HBLs in
$\alpha_{rx}$ versus $\alpha_{ro}$ diagram is different from that of FSRQs
and LBLs. This indicates that HBLs have different spectral properties
from FSRQs and LBLs. The results are consistent with the reported results of
\citet{Xie04a}, who found that HBLs and LBLs locate in different
regions in the $\alpha_{ox}-\alpha_{x\gamma}$ plane, but LBLs and FSRQs
occupy the same region in $\alpha_{ox}-\alpha_{x\gamma}$ plane. In addition, our results also agree with that reported by \citet{Fan12} and \citet{Lyu14}.

\begin{figure}
\includegraphics  [width=5in, angle=0]{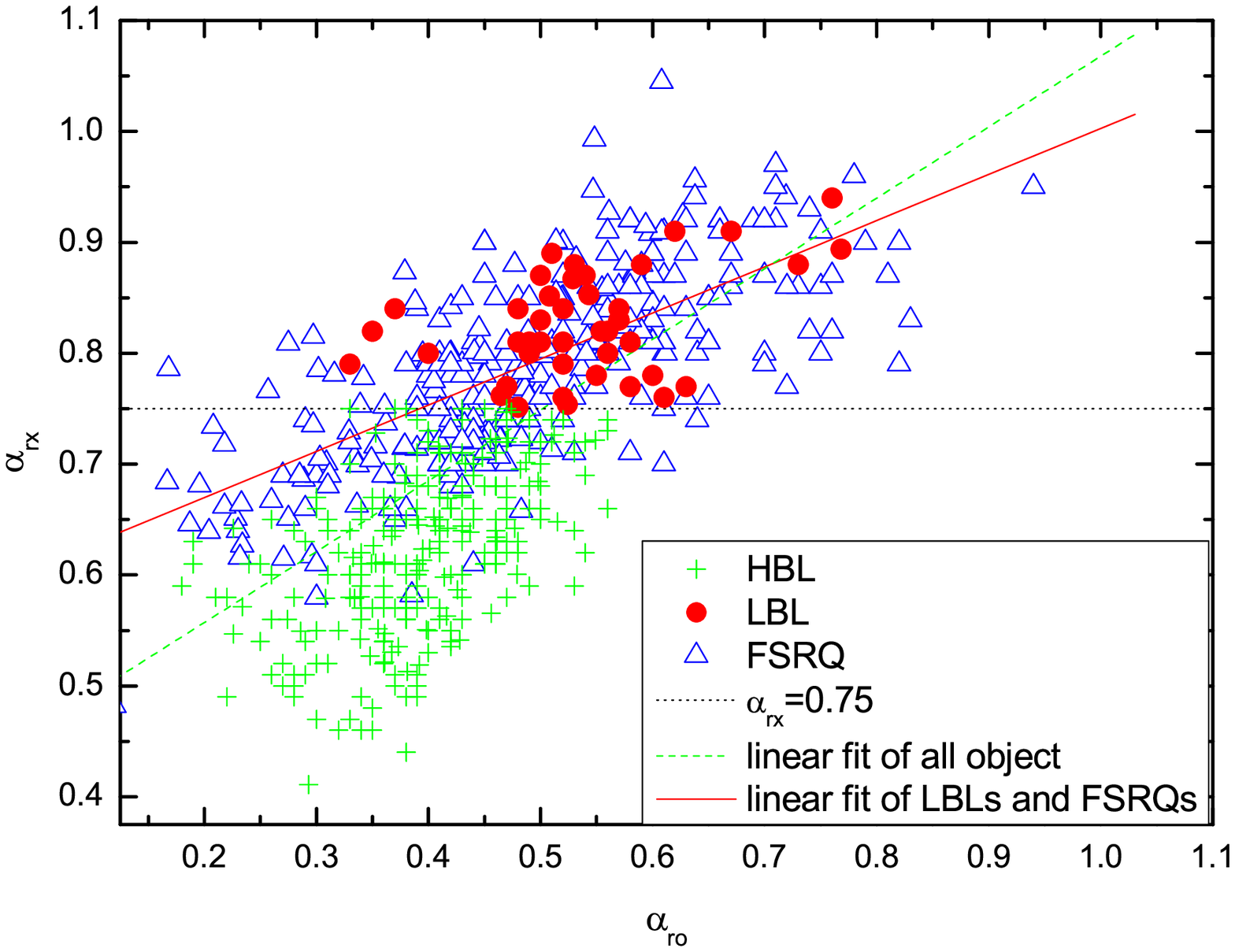}
 \caption{The relationship between the broadband spectral indices $\alpha_{ro}$ and $\alpha_{rx}$ for the sources in our sample.
 \label{fig4}}
\end{figure}

\subsection{Diagram of $\alpha_{rx}$ versus $\alpha_{ox}$}

In Figure 5, $\alpha_{rx}$ versus $\alpha_{ox}$ is plotted for our
SDSS blazar sample. We can find that the distribution of three kinds
of blazars revealed from Figure 5 is similar to that of Figure 4. For the whole sample, Figure 5 shows a significant correlation between $\alpha_{rx}$
and $\alpha_{ox}$, and the linear regression analysis yields
\begin{equation}
\alpha_{rx}=(0.30\pm 0.01)\alpha_{ox}+(0.36\pm0.02),
\label{eq:Lebseque3}
\end{equation}
with a correlation coefficient $r=0.62$ and a chance
probability $p<10^{-4}$. Moreover, Figure 5 shows that there is a weak correlation between $\alpha_{rx}$
and $\alpha_{ox}$ for the FSRQs and LBLs sample, and the linear
regression analysis equation is

\begin{equation}
\alpha_{rx}=(0.07\pm 0.02)\alpha_{ox}+(0.71\pm0.03),
\label{eq:Lebseque4}
\end{equation}
with a correlation coefficient $r=0.17$ and a chance
probability $p=0.0016$. In Figure 5, one can note that the
majority of FSRQs and LBLs also occupy the same region, but HBLs
occupy a separate distinct region, which is also consistent with
previous results.

\begin{figure}
\includegraphics  [width=5in, angle=0]{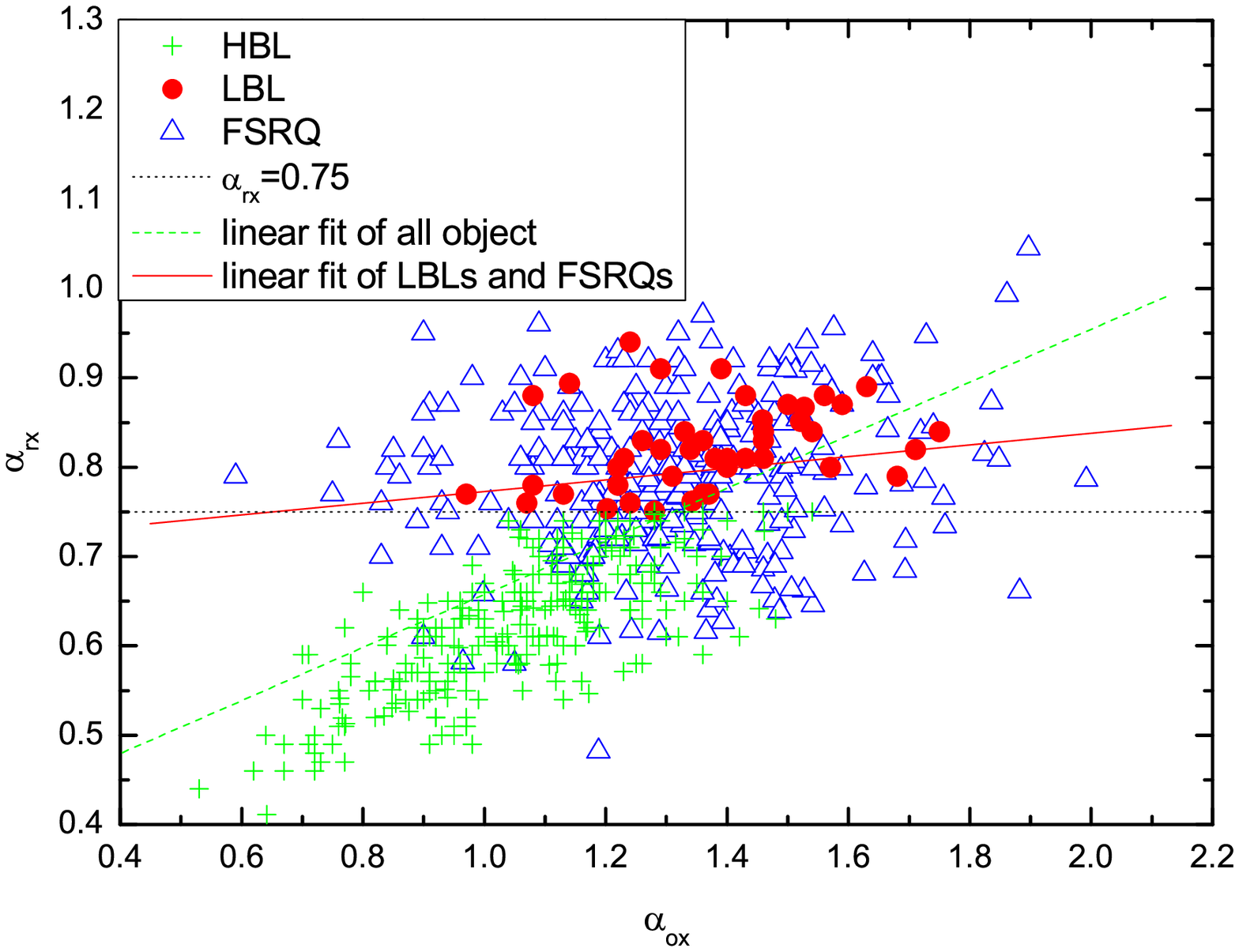}
 \caption{The relationship between the broadband spectral indices $\alpha_{ox}$ and $\alpha_{rx}$ for the sources in our sample.
 \label{fig5}}
\end{figure}

\subsection{Diagram of $\alpha_{ox}$ versus $\alpha_{ro}$}
Based on the broad band spectral index $\alpha_{ox}$ versus $\alpha_{ro}$, we investigated the relationships between $\alpha_{ox}$ and $\alpha_{ro}$. Figure 6 plots $\alpha_{ox}$ versus $\alpha_{ro}$. A linear
regression analysis shows that there is a weak or even no correlation between
$\alpha_{ox}$ and $\alpha_{ro}$ ($r=-0.11$ and $p=0.054$) for whole sample. This
independent relation is obviously inconsistent with the foregoing
correlation revealed from Figure 4 and Figure 5. However, the correlation
is significant when considering FSRQs and LBLs sample and the the linear
regression analysis equation is
\begin{equation}
\alpha_{ox}=(0.87\pm 0.07)\alpha_{ro}+(1.75\pm0.04),
\label{eq:Lebseque6}
\end{equation}
where the correlation coefficient is $r=-0.56$ and the chance
probability is $p<10^{-4}$. This suggests that HBLs are different from FSRQs, but LBLs are similar to FSRQs.
 In addition, Figure 6 also shows that most of the FSRQs and LBLs
locate the same region in $\alpha_{ox}$ -$\alpha_{ro}$ plot, but
HBLs occupy a separate distinct region in $\alpha_{ox}$ versus
$\alpha_{ro}$ plane. This is consistent with the distribution shown in Figure 4 and Figure 5. This supports the foregoing results: FSRQs and LBLs have similar
spectral properties, but HBLs have distinct spectral properties.



\begin{figure}
\includegraphics  [width=5in, angle=0]{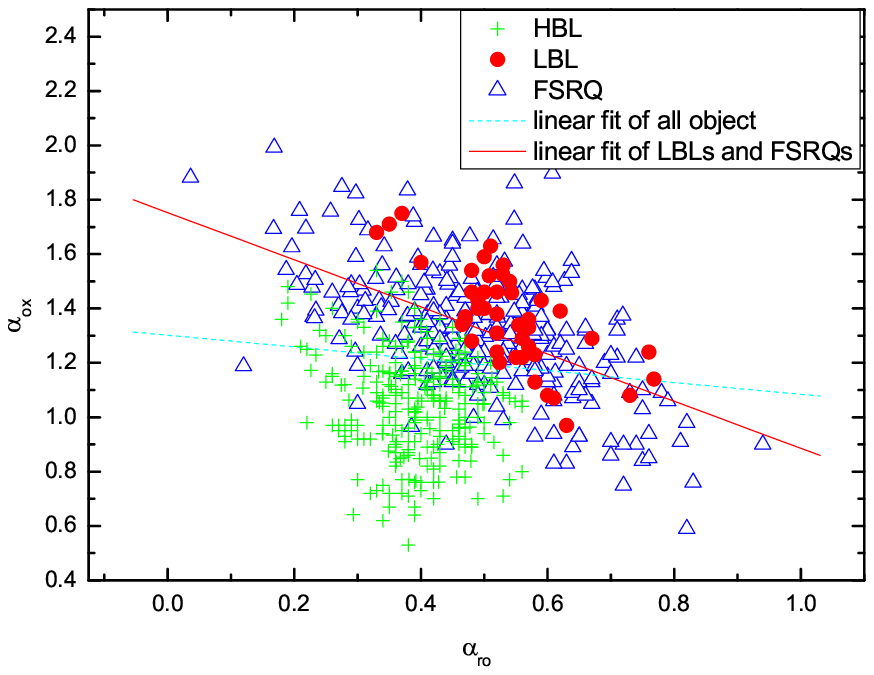}
 \caption{The relationship between the broadband spectral indices $\alpha_{ro}$ and $\alpha_{ox}$ for the sources in our sample.
 \label{fig6}}
\end{figure}

\subsection{Summary}
As noted above, from Figure 4 and 5, there is a strong correlation
between $\alpha_{rx}$ and $\alpha_{ro}$, and as well as between
$\alpha_{rx}$ versus $\alpha_{ox}$ for the whole sample, which provides some more
evidence for the conclusion reported
by \citet{Foss98} and \citet{Ghis98}. Namely, there is a unified
scheme for blazars. On the other hand, from Figure 4, 5, and 6, one
can note that there are also some different distributions from the
blazar sequence reported by \citet{Foss98} and \citet{Ghis98}. In
the color-color diagram, HBLs and FSRQs occupy seperated regions,
while the LBLs and FSRQs mix together, which is consistent with that
reported in some
literatures \citep[e.g.][]{Chen13,Fan12,Li10,Lyu14,Xie04a}. This suggests that FSRQs and LBLs have a
similar property, but HBLs have a distinct property.

\section{Discussion and Conclusions}
Based on the Slew survey, the 1-Jy samples of BL Lacs and the 2-Jy
sample of FSRQs, \citet{Foss98} studied the systematics of the SEDs of blazars using data
from the radio to the $\gamma$-ray band, and found that three
different kinds of blazars follow an almost continuous spectral
sequence: from FSRQs through LBLs to HBLs. \citet{Ghis08} revisited the so called "blazar sequence", and proposed that the power of the jet and the SED of its emission are linked to the two main parameters of the accretion process. This similar trend was also
obtained by other authors
\citep[e.g.][]{Xie04a,Samb96,Ghis98,Bott02,Mara08} who found that similar
physical processes operate in three kinds of blazars.
However, \citet{Ant05}
found that there are selection effects for the
"blazars sequence". Moreover, some authors found that HBLs do not
follow the blazars sequence \citep[e.g.][]{Chen13,Fan12,Giom05,Li10,Pado03,Pado07} .

In this paper, we computed the distributions of the radio (at 5
GHz), optical (at 5000 ${\AA}$), and X-ray (at 1 keV) luminosities.
The luminosities  distributions reveal that it is an important
parameter to distinguish FSRQs and BL Lacs objects, and the
distributions are continuous for three kinds of blazars. The
luminosities of FSRQs are lager than that of BL Lacs objects, which
is in good agreement with the arguments reported by other authors
\citep{Abdo09a,Foss98,Ghis98,Samb96}. The distributions of radio
luminosity support the blazars sequence reported by \citet{Foss98}
and \citet{Ghis98}. However, the distributions of optical and X-ray
luminosities do not support the sequence.

The broadband energy distribution shows that three kinds of blazars have
different spectral properties. It also shows that most FSRQs and LBLs mix together in the color-color
diagram (see Figure 4, 5, and 6), which is consistent with previous
results \citep{Fan12,Li10,Samb96,Xie04a,Zhen07}. This suggests that they
have similar spectral properties, which provides some more evidence for the conclusion of unified
scheme. However, the location of HBLs is separate with that of FSRQs
and LBLs in the color-color diagram, which reveals that HBLs have
different SEDs from FSRQs and LBLs. This suggests that the results
from SDSS sample do not support the so called "blazar sequence",
which is consistent with the results reported by the other authors
\citep{Ant05,Fan12,Chen13,Li10,Pado03,Pado07,Zhan12}. The spectral
sequence obtained by \citet{Foss98} may be related to the selection
effects, because the sample used by \citet{Foss98} are classic, high
flux limit surveys in the radio and X-ray \citep{Ant05}. Our sample
used in the paper is a large sample including 606 blazars, which
would result in a unbiased view of blazar spectral properties.
Moreover, Figure 7 gives the relations between the redshift and spectral
indices. Figure 7 shows that the spectral indices is independent of
redshift, which suggests that the selection effects of our sample
are weak.


\citet{Ghis98} suggested the level of cooling is different for different subclasses of blazars. FSRQs suffer stronger cooling, and synchrotron emission peaks at much lower frequency. However, the cooling is less important for HBLs, and the energetic particles can produce synchrotron and inverse Compton (IC) emission up to high frequency. The level of cooling of FSRQs stronger than the one of HBLs may be due to the external radiation field \citep{Ghis98}. \citet{Geor01} suggested that the radiating jet plasma in weak sources is outside the broad line scattering region (BLR), while it is within in the power source. This implies that the location of emitting region between HBLs and FSRQs might be very different \citep{Cost09}. The jet energy of FSRQs would dissipate within the BLR, leading that the high energy electrons in the jet will suffer greater cooling \citep{Chen11}. \citet{Ghis10} suggested that the $\gamma$-ray emission from FSRQs is likely from the Compton scattering of an external radiation source, while for HBLs SSC is able to provide a good fit to the $\gamma$-ray emission. In addition, based on the physical properties of relativistic jets, \citet{Yan14} found that the one-zone SSC model can successfully fit the SEDs of HBLs, but fails to explain the SEDs of LBLs. Moreover, they also suggest that the ratios of the Compton to the synchrotron peak energy fluxes of LBLs are greater than those of HBLs and IBLs, and then LBLs are Compton dominated \citep{Yan14}. This suggests that there is an external radiation field for LBLs. Therefore, the levels of cooling of FSRQs and LBLs is stronger than HBLs, which lead that the synchrotron emission peaks of FSRQs and LBLs is lower than ones of HBLs. \citet{Abdo10} found FSRQs and LBLs are the low synchrotron peaked
blazars, while HBLs are high synchrotron peaked
blazars. \citet{Fan12} suggest that if the synchrotron peak frequency moves to the lower frequency, then the IC peak frequency may also move to the lower frequency. Thus, the X-ray of LBLs and FSRQs are from the combination of synchrotron emission and the IC emission, while the X-ray of HBLs are from the
synchrotron emission of very high energy electrons \citep{Abdo10,Fan12}.
In addition, the different SEDs between HBLs, FSRQs, and LBLs may be
related to the different intrinsical environments around the
blazar's nucleus. The intrinsical environments of FSRQs and HBLs have
a clear, physics difference: the environment of HBLs is "cleaner"
than that of FSRQs \citep{Cost09}. The central regions of FSRQs are rich in gas and dust, which would lead to a high accretion rate onto the central supermassive black hole \citep{Bott02}. Moreover, the material would efficiently
reprocess and scatter the accretion disk radiation. This would
lead to the observed strong optical emission lines in the
BLR and to a high energy density of the
external soft-photon field in the jet \citep{Bott02}.

Although HBLs have different SEDs from FSRQs and LBLs, the
significant correlation revealed from Figure 4 and 5 suggests that
there is a unified scheme for whole sample, which is consistent with the previous conclusion
reported by other authors
\citep{Coma97,Foss98,Ghis98,Li10,Samb96,Xie01,Xie04a}. This hints
that there is a similar physical processes operating in all objects.
In the case of the blazar-type sources where the emission is usually
associated with a stream of relativistic jet, the overall spectrum
is determined by the energy spectrum of the electrons as well as by
the variation of the physics quantities along the jet
\citep{Bege84}. HBLs, LBLs, and FSRQs have a significant correlation
in the color-color diagram (see Figure 4 and 5), which implies that
similar physical processes operate in all objects under a range of
intrinsic physics conditions or beaming parameter. On the other
hand, the difference among three subclasses of blazars, revealed
from color-color diagram (see Figure 4, 5, and 6), should be
attributed to the different level of cooling and the intrinsical different environments around the blazars
nucleus for different subclasses blazars, which lead to
different optical and X-ray spectra for different kinds of blazars.

\begin{figure}
 \includegraphics [width=5.5in, angle=0]{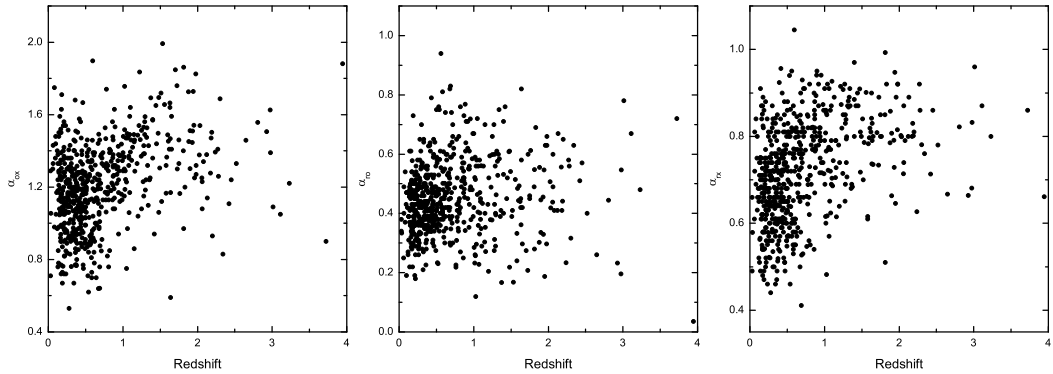}
 \caption{Relations between redshift and spectral indices.
 \label{fig7}}
\end{figure}

\normalem
\begin{acknowledgements}
We are grateful to the anonymous referee for useful comments. We are grateful for the help from Liang Chen. This research has made use of the SDSS database.
This work
is supported by the National Natural Science Foundation of China
(10878013), and the Natural Science Foundation of Yunnan Province
(2011FZ081, 2012FD055, 2013FB063), and the Program for Innovative
Research Team (in Science and Technology) in University of Yunnan
Province (IRTSTYN), and Science Research Foundation of Yunnan
Education Department of China (2012Y316), and the Young Teachers
Program of Yuxi Nurmal University. In addition, the work of Yunguo Jiang is supported by the NNSFC with Number 11403015.
\end{acknowledgements}


\label{lastpage}
\end{document}